\documentclass[11pt,a4paper]{article}
\usepackage{psfig}
\usepackage{a4}
\usepackage{amsfonts}
\usepackage{isolatin1}
\newcommand\BE{\begin{equation}}
\newcommand\EE{\end{equation}}
\newcommand\BEA{\begin{eqnarray}}
\newcommand\EEA{\end{eqnarray}}
\begin{document}
\title{ 
Temporal Structures in Shell Models.\\}
\author{Fridolin Okkels \\
{\it Dept. of Optics and Fluid Dynamics, Risø National Laboratory}\\
and {\it Centre for Chaos and Turbulence Studies},\\
{\it The Niels Bohr Institute, University of Copenhagen}
}
\maketitle
\begin{abstract}
The intermittent dynamics of the turbulent GOY shell-model is characterised by 
a single type of burst-like structure, which moves through the shells 
like a front. This temporal structure is described by the dynamics of 
the instantaneous configuration of the shell-amplitudes revealing a 
approximative chaotic attractor of the dynamics. 
\end{abstract}
\section{Introduction}
One of the main goals in current turbulence research is to understand the 
effect of intermittency in turbulence. It has long been known that 
intermittency produces corrections to 
the classical Kolmogorov $-5/3$ scaling law and to other moments of the 
energy spectrum in the inertial range \cite{Fri1,Kol1}. 
Still very little is known about the intense intermittent structures found 
in turbulent flows \cite{Tab1}.\\
\indent
Over the last decade turbulent shell-models have been studied intensively
because of their simplicity and excellent agreement of their statistics in 
comparison with that of experimentally measured turbulence 
\cite{Gle1,OY1,Jen1,Kad1,Gre1,Bif1}. 
In a way these models reproduce the statistics far better than expected from
their simplicity, so the general idea has been to reveal
the nature of the dynamics of these models and then afterwards relate this
experiences to the full problem of turbulence.\\
\indent
While the main approach to shell models has been statistical, much can be 
learned from the study of the temporal structures naturally arising in the
model. Because of the intense dynamics during these temporal events they 
are called bursts. 
Only recently have these temporal structures been thoroughly 
studied together with the nature of their creation \cite{Dom1,Okk1}.
Among the large numbers and types of different shell-models the present
work is based on the successful GOY shell-model \cite{Gle1,OY1}.\\
\indent
The paper is organised as follows: The first part gives a detailed description 
of the bursts of the standard GOY-model. The second part shows that an 
approximative chaotic attractor of the model exists expressed by the 
collective dynamics of the neighbouring shells.
\section{The GOY-model}
All shell-models simulate the flow of energy through wavenumber space
in fully developed turbulence.
The models consist of a system of coupled ordinary differential equations 
where the energy is injected into the low wave-numbers by a constant 
forcing term. The energy then cascades up to the high wave-numbers by means of 
a coupling term where it is dissipated away by a viscosity term.
\subsection{Construction}
In the GOY model wave-number space is divided 
into $N$ separate shells with characteristic wavenumbers 
$k_n = k_0 \lambda^n \ \ (\lambda=2)$ where $n=1,\ldots ,N$ and $k_0$ is 
a constant determining the smallest wavenumber in the model. Each shell is 
assigned a complex amplitude $u_n$ which can be imagined as the velocity 
difference on a scale $\ell_n=1/k_n$.
By assuming conservation of phase space, energy and helicity
and interactions among the nearest and next nearest
neighbour shells, one can arrive at the following set of evolution equations 
\cite{OY1,Kad1,Bif1}
\BE
\label{Gen-Un}
\left({d\over dt}+\nu k_n^2\right)u_n = ik_n(u_{n+1}^*u_{n+2}^*
-\frac{\delta}{2}u_{n-1}^*u_{n+1}^*-\frac{1-\delta}{4}u_{n-2}^*u_{n-1}^*)+
f\delta_{n,4}
\EE
with boundary conditions $u_{-1}=u_0=u_{N+1}=u_{N+2}=0$, and constant forcing 
$f$ on the fourth shell.\\
\indent
The set (\ref{Gen-Un}) of $N$ coupled ordinary differential equations is
numerically integrated by standard techniques \cite{Okk1}.  
In the simulations, we use the 
following standard values: $\delta =1/2,\ N=19,\ \nu = 10^{-6},\ k_0 = 2^{-4},
\ f = (1+i)*0.005$ as found in earlier work \cite{Gre1,Jen1,Kad1,Bif1}.
\subsection{Conservation laws, fixed-points and invariance}
The strength of the shell-models relates to their simplicity in construction, 
which can be justified by imposing the same conservation-laws 
and invariants as the Naiver-Stokes Eq.~i.e.~the basic equations 
governing fluid dynamics.
As for other shell-models the GOY-model exhibits these conservation laws in the
absence of forcing and viscosity $(f=\nu=0)$ reducing the right side of 
Eq.\ref{Gen-Un} to the coupling-term. \\
\indent
The conservation of phase-space is enforced as the 
coupling term does not contain $u_n$. The remaining conserved 
quantities are quadratic i.e.~they can be written in the form: 
$Q_\alpha = \sum k_n^\alpha |u_n|^2$. Using the relation 
$\frac{1}{2}\dot{(u_n^2)} = u_n\dot{u}_n$
and inserting $Q_\alpha$ in the model, the coupling term becomes three terms of
three successive amplitudes multiplied together. Comparing these three terms 
gives the following relation on 
$\alpha$:
$1-\delta 2^{\alpha}-(1-\delta)2^{2\alpha} = 0$ with two solutions:
$\alpha=0$ and $\alpha = -\ln_\lambda (\delta-1)$.
The first ($\alpha=0$) corresponds to the conservation of energy while the 
other solution corresponds to helicity conservation in the case of three 
dimensional turbulence $(\delta<1)$ and to enstrophy conservation in the case 
of two dimensional turbulence $(\delta>1)$ \cite{Kad1,Gre1,Okk1}.\\
\indent
The fact that the coupling term multiplied by $u_n$ gives three terms all 
similar within pre-factors and a displacement in indices makes the dynamics 
of the model invariant to the following changes in the complex phase:
\BEA
\label{eq.invar}
u_n     & \to & e^{i\,\alpha}\,u_n   , \nonumber \\
u_{n+1} & \to & e^{i\,\beta-\alpha}\,u_{n+1} , 
\qquad \textrm{where $n$ modulus } 3=1 \\
u_{n+2} & \to & e^{i\,-\beta}\,u_{n+2} , \nonumber 
\EEA 
and $\alpha$ and $\beta$ are free parameters. This invariance affects not 
only the phases but also the dynamics of the model since every third shell 
tend to follow the same behaviour.\\
\indent
Thinking of the GOY-model as a dynamical system a basic thing to 
study is the fixed-points of the model: $\dot{u}_n = 0,\ n=1\ldots N$.
Requiring again an inviscid and unforced model $(f=\nu=0)$ gives two 
non-trivial scaling fixed-points: $u_n = k_n^{-z}\,g(n)$ with 
$z=\frac{1}{3}$ and $z=\frac{1}{3}(1-\ln_\lambda(\delta-1))$ where $g(n)$
is an arbitrary function of period three in $n$ coming from the invariance of 
the model. The first fixed-point: $u_n \sim k_n^{-1/3}\,g(n)$ 
corresponds to the Kolmogorov $-\frac{5}{3}$ scaling-law and will be called the
Kolmogorov fixed-point while the other solution result in an alternation of 
the amplitudes \cite{Gre1,Okk1}. In spite the simplicity of the these 
fixed-points they play a crucial role in the later analysis of the model.
\section{Dynamics of the model}
\begin{figure}[ht]
\centerline{\psfig{file=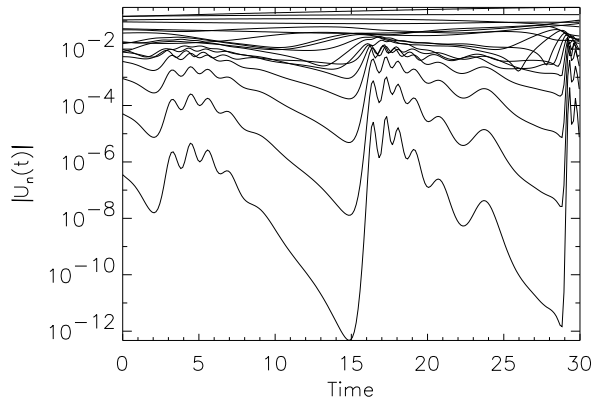,height=6cm}}
\caption{A typical evolution of the norm of the amplitudes 
$|u_1|\ldots|u_{18}|$}
\label{fig.Norm}
\end{figure}

At large time-scales the dynamics of the model may seem stochastic but as 
the time-span refines distinct spikes emerge and
in the end the dynamics is noiseless and fully resolved even during the most
dramatic changes. To observe the general behaviour we monitor $|u_n(t)|$ 
as a function of time and because of large variations in magnitude it is 
shown in a semi-logarithmic plot in Fig.\ref{fig.Norm}.
The higher shells has the lowest absolute value and fastest variations 
while the lower shells vary over a longer time-span.

Two main features strike from Fig.\ref{fig.Norm}: All the higher shells 
evolve in a synchronised manner and the evolution follows a general pattern of 
strong bursts interchanged by oscillatory relaxation. Bursts occur randomly in 
time with great variations in their strength.
\subsection{Organisation of shell-dynamics}
\begin{figure}[htb]
\centerline{\hbox{
\psfig{file=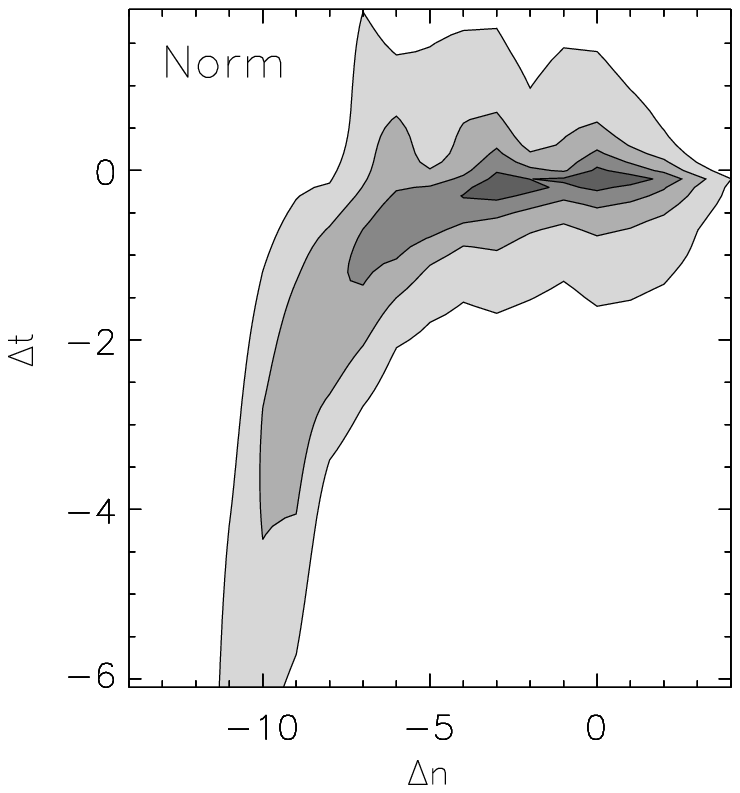,height=7cm}
\psfig{file=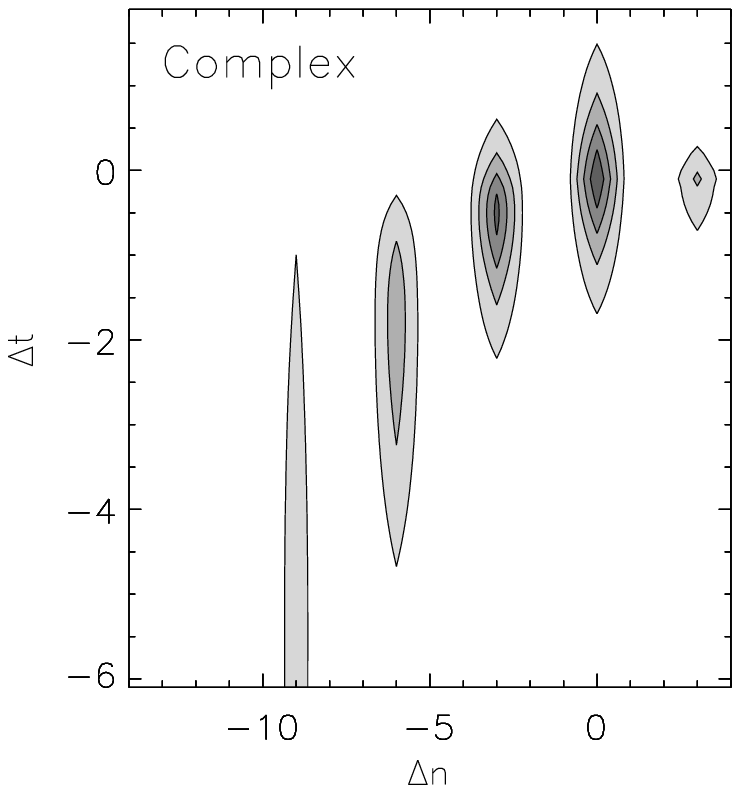,height=7cm}
}}
\caption{The two-point correlation in shells and time first for the norm of 
the amplitudes and second for the pure amplitudes, based at the 15'th shell.}
\label{fig.NT-Corr}
\end{figure}

The synchronisation of the different shells is a result of the coupling 
between the shells making the model self-organise into the 
types of behaviour seen in Fig.\ref{fig.Norm}.
The organisation in the model is showed most clearly by the 
local two-point correlation function measuring how the dynamics of a 
given shell is correlated to its neighbourhood of both shells and in time.
It is defined using the following two shortcuts
\[ U_0=U_{n_0}(t), \quad U_{\Delta}=U_{n_0+\Delta n}(t+\Delta t)\]
by:
\BE
\Gamma(\Delta t, \Delta n) = C(U_0,U_{\Delta})
= \frac{\overline{U^*_0\cdot U_{\Delta}}-
\overline{U^*_0}\cdot \overline{U_{\Delta}}}
{\sqrt{\left(\overline{|U^2_0|}-\overline{|U_0|}^2\right)
\left(\overline{|U^2_{\Delta}|}
-\overline{|U_{\Delta}|}^2\right)}} 
\label{eq.NT-Corr}
\EE
and where the averages are taken over time.\\
\indent
The information gained from the $|\Gamma(\Delta t, \Delta n)|$ is divided into 
two parts: First only the norm of the complex amplitudes is correlated 
replacing $U_0$ and $U_{\Delta}$ by their norms. This is shown in the left 
part of Fig.\ref{fig.NT-Corr} as a contour-plot with dark as the strongest 
normalised correlation. Second the full complex amplitudes are correlated and 
showed in the same manner in the right part of Fig.\ref{fig.NT-Corr}. 
Both correlations have $n_0=15$ and averaged over $40.000\ n.u.$ which 
correspond roughly to a time-span of approximately $4000$ successive bursts.

The left plot show that all the amplitudes in the model is strongly
correlated from the forcing at the $4^{\textrm{th}}$ shell up to the highest 
shells. 
This strong correlation is due to the organisation
of the amplitude dynamics during both bursts and the succeeding strong
oscillations. The same plot also shows the motion of the burst through the
shells by the time-shift of the correlation-peeks for increasing $\Delta n$.
When taking the amplitude phases into account the correlation function
changes radically as seen in the right plot. Now only every third amplitude
are correlated and this comes as a result of the period three invariance of
the model. 
This plot shows also how the characteristic time-scale changes among
the different shells. It is seen by the extent of the
correlation-peeks in time which decreases with shell-number.
When relating the characteristic time-scale to the 
turnover-time ($\tau_n$) this dependence comes direct from dimension 
analysis \cite{Fri1}.
\subsection{Front motion during burst}
\begin{figure}[ht]
\centerline{\psfig{file=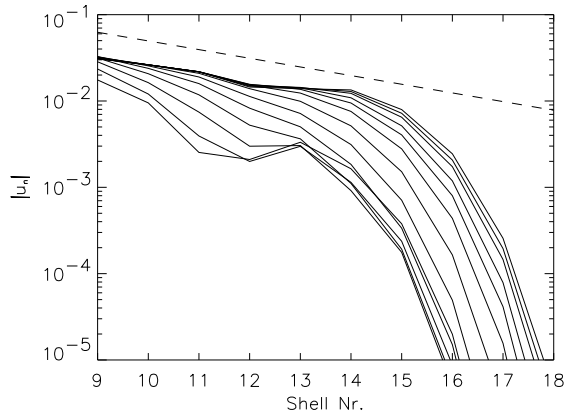,height=6cm}}
\caption{Snapshots of $\log|u_n|$ during the {\it cascade} of a burst}
\label{fig.front}
\end{figure}

The motion of the bursts 
is a part of a more general motion of different 
organisations of the amplitudes travelling with exponential increasing 
speed from the lower towards the higher shells where they vanish because of
viscosity \cite{Dom1}. 
A way to see this is to look at the changes in the instantaneous amplitude
spectre during the motion of a burst. This is shown in Fig.\ref{fig.front} by 
snapshots of $\ln|u_n(t)|$ vs. $n$ where the time between snapshots decrease 
by a factor of $1/\sqrt{2}$ giving roughly an equidistant motion of the burst.
As for all other bursts Fig.\ref{fig.front} reveals that the burst travels 
through the shells as a front keeping the same overall shape.
Just at the maximum rise of the amplitudes the overall scaling exponent of the
inertial range is a bit lower than the Kolmogorov scaling-law shown by the
dashed line in Fig.\ref{fig.front}. Immediately after the last snapshot all 
the shells enters the oscillatory state.
\subsection{Real-valued model}
Due to the invariance in the model the creation and behaviour of the bursts 
are unaffected by the complex phase of the amplitudes. A model in terms of 
real values will therefore be used in the following analysis:
\BE
\label{Real-GOY}
\left({d\over dt}+\nu k_n^2\right)r_n = -k_n(r_{n+1}r_{n+2}
-\frac{1}{4}r_{n-1}r_{n+1}-\frac{1}{8}r_{n-2}r_{n-1})+f\delta_{n,4}
\label{eq.real-GOY}
\EE
having $r_n=|u_n|$, no conjugations and ``$-1$'' instead of ``$i$'' in front 
of the coupling-term \cite{Okk1}.
\section{Local variables}
\begin{figure}[ht]
\centerline{\psfig{file=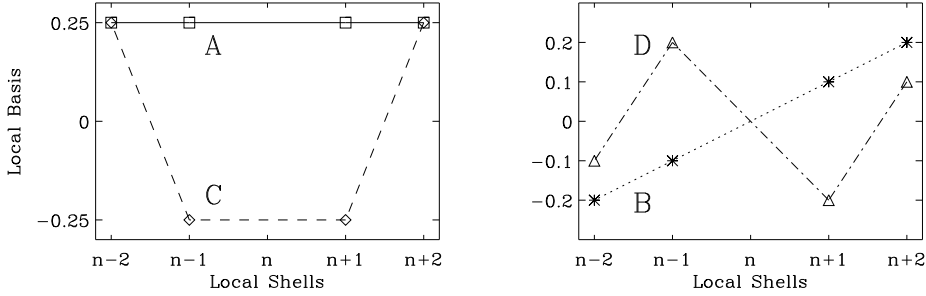,height=4.5cm}}
\caption{The basis of the local variables}
\label{fig.LV-basis}
\end{figure}

From the construction of the model the dynamics of a given shell depends only 
on the instantaneous configuration of the neighbouring shells and it has
 no explicit dependence on the present or past states.
If we at first restrict ourself to the inertial range neglecting forcing and 
viscosity the neighbouring shells may be seen as {\em a local phase-space} of 
a shell since their configuration through the coupling-term exactly 
determines the instantaneous dynamics $(\dot{r}_n)$ of the amplitude
$r_n$.
To characterise this local phase-space each set of neighbouring shells will be 
called the 
{\em local shells}: $\vec{L}_n=(r_{n-2},r_{n-1},r_{n+1},r_{n+2})$ of the 
n$^{\textrm{th}}$ shell, and they should not be seen as part of the other 
amplitudes but rather as an isolated set of variables determining 
$\dot{r}_n$.\\
\indent
The configuration of $\vec{L}_n$ will be described by first choosing the 
slope of $\ln_{\lambda}(\vec{L}_n)$ which is nothing but the local scaling 
exponent at the n$^{\textrm{th}}$ shell. To continue we define 
\BE
\vec{\eta}_n \equiv \ln_{\lambda}(\vec{L}_n)
\EE
and choose the mean, curvature and third order component of 
$\vec{\eta}_n$. 
This gives the {\em local variables}: $\vec{P}_n=(A_n,\,B_n,\,C_n,\,D_n)$ of 
$r_n$ defined as the coefficients of the projection of $\vec{\eta}_n$ on the 
orthogonal basis given by the matrix $\mathbf{T}$ :
\BE
\vec{ \eta }_n=\mathbf{T} \cdot \vec{P}_n ,\quad \vec{L}_n=2^{\vec{\eta}_n} 
\label{eq.Local-Var}
\EE
where 
\BE
\mathbf{T}=\left( \begin{array}{cccc}
\alpha & 2\beta  & -\alpha & \beta \\ 
\alpha & \beta  & \alpha  & -2\beta\\
\alpha & -\beta & \alpha  & 2\beta \\
\alpha & -2\beta & -\alpha & -\beta \end{array} \right)
\label{eq.T-Matrix}
\EE
and $\alpha=1/4$ and $\beta=1/10$.\\
\indent
The basis of the local variables is plotted in Fig.\ref{fig.LV-basis} 
showing how it can be characterised as a simple ``Taylor-series'' expansion
of $\vec{\eta}_n$. These variables are believed to be the right variables to 
monitor the dynamics 
of the model since they describe globally the configuration of the local 
shells instead of focusing on the individual neighbouring shells.
The local scaling of shell-models has been studied earlier \cite{Bif1,Gre1},
but this was the instantaneous local scaling averaged over all shells and 
using a coarse-grained time resolution.
\subsection{Application to the model}
\begin{figure}[ht]
\centerline{\psfig{file=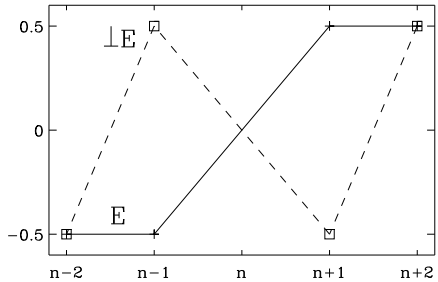,height=4.5cm}}
\caption{The period three invariance of the local basis}
\label{fig.P3-invar}
\end{figure}

To implement Eq.\ref{eq.Local-Var},\ref{eq.T-Matrix} into the model we assume 
the components of $\vec{\eta}_n$ to behave smoothly in $n$ such that 
$r_n \approx 2^{A_n}$, giving:
\BE
\dot{r}_n=-\, k_n \, 2^{2A_n}\left(2^{3B_n-D_n}
-\mbox{$\delta \over 2$}2^{-2C_n}
-\mbox{${1-\delta} \over 4$} 2^{-3B_n+D_n} \right) -\nu \, k_n^2 \, 2^{A_n}.
\label{eq.Local-Var-GOY}
\EE
Eq.\ref{eq.Local-Var-GOY} gives direct evidence of the period three 
invariance of the model: Since the dynamics only depends on 
the combinations ($3B_n-D_n,\,C_n,\,A_n$), we define $E_n \equiv 3B_n-D_n$. 
The model is then invariant to the orthogonal 
component of $E_n$: $\perp \! E_n = 3D_n+B_n$ which is nothing but the period
three invariance as seen in in Fig.\ref{fig.P3-invar}.

From the construction of Eq.\ref{eq.Local-Var-GOY} is should be noted that the 
sign of
$\dot{r}_n$ and thereby the monotony of the dynamics is only a function of
$E_n$ and $C_n$ when neglecting the viscosity-term. Because $A_n$ is 
outside the brackets it will affect the response-time of the dynamics.
Now the dynamics of the n$^{\textrm{th}}$ amplitude is only determined by 
three local variables: 
\[\vec{V}_n=(E_n,\,C_n,\,A_n).\]
Even though this new set of local variables ($\vec{V}_n$) form a efficient 
phase-space it should not be confused with the actual 
2N-dimensional phase-space of the free variables in the model.
\subsection{The Local Attractor of the model}
Since $\vec{V}_n$ is a local phase-space the trajectory of $\vec{V}_n(t)$ 
in time will describe a {\it three dimensional local attractor} of the 
n$^{\textrm{th}}$ shell dynamics.
Fig.\ref{fig.LV-Traj+Surf2} show the local attractor of the 
14$^{\textrm{th}}$ shell during a time-span of two successive 
bursts where some additional features is placed to explain the dynamics of 
the attractor: 
\begin{figure}[ht]
\centerline{\psfig{file=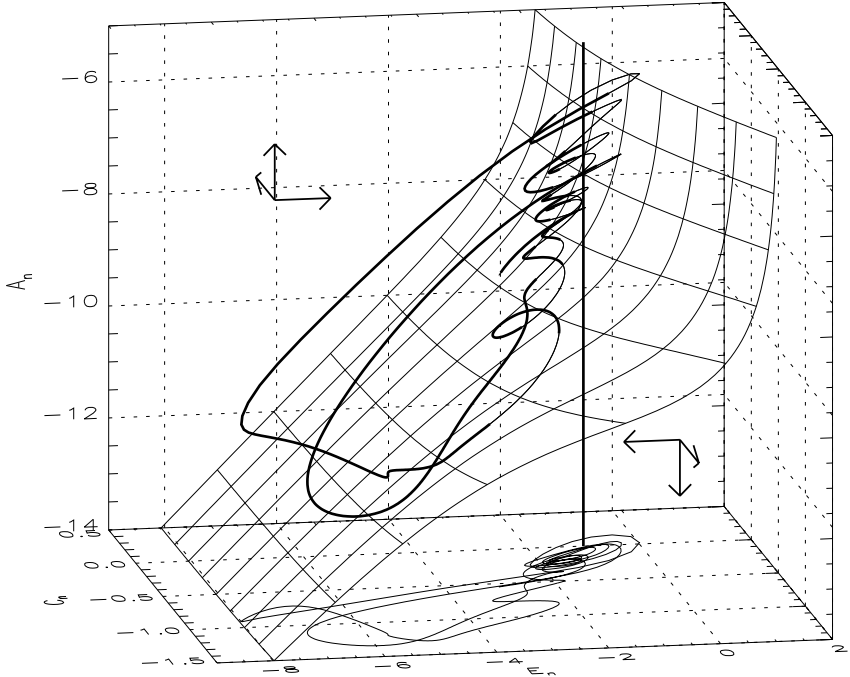,height=9.5cm}}
\caption{The local attractor of the 14'th shell and its projection on 
a $(E_n,C_n)$-plane together with the surface of $\dot{r}_n=0$, the 
Kolmogorov fixed-point-line and arrows of characteristic flow.}
\label{fig.LV-Traj+Surf2}
\end{figure}

First we note that the trajectory is projected down on a $(E_n,C_n)$-plane to 
help giving a three dimensional understanding of the attractor. Then 
we focus on the vertical line which correspond to the Kolmogorov 
fixed-point given by $(E_n,C_n,A_N)=(-1,0,\cdot)$. Right after every 
bursts the trajectories encircles this line during the
relaxations.
As the oscillations die out the dynamics slow down making the trajectories 
stay close to the region of $\dot{r}_n \approx 0$ in $\vec{V}_n$. 
In Fig.\ref{fig.LV-Traj+Surf2} the curved sheet is the manifold of 
$\dot{r}_n = 0$ derived from Eq.\ref{eq.Local-Var-GOY} and it is seen how the 
trajectory stays close to the manifold. ( note that the 
trajectory is shown thinner for negative $\dot{r}_n$)\\ 
\indent 
When a burst approaches from the lower shells it affects the configuration of
local shells forcing the trajectory away from the manifold. This causes
$\dot{r}_n$ and thereby $r_n$ to increase rapidly making the shell participate
the burst.
During the burst the trajectory approaches the Kolmogorov fixed-point-line
 around which it begins to circle again etc. The same behaviour repeats 
throughout the evolution of the model {\em making the local attractor capture 
all the general dynamics of the model}.\\
\indent
Every other shell participating in the burst has qualitatively the same local 
attractor with the same characteristics.
It should be noted that if the viscous term only affects the last shells, 
abandoning the inertial range, the model would still produce bursts and in 
this case the oscillations would not bend off but follow the Kolmogorov 
fixed-point strait down until the next burst approaches.
\section{The cause of intermittency}
From the behaviour of the local attractor it is possible to explain the 
intermittent shift between bursts and oscillatory relaxation of the model. 
What is needed is the answers to the following two questions: 
Why is the manifold of $\dot{r}_n=0$ stable, attracting the oscillatory state 
into a relaxing period and what changes this stability as a burst approaches.
\subsection{Creation of the relaxing period.}
To analyse the stability of the manifold we have to know the flow in the
phase-space $\vec{V}_n$ and this will be done by estimating  
$\dot{A}_n,\dot{E}_n,\dot{C}_n$:\\
\indent
First we assume again $r_n \approx 2^{A_n}$ to get 
$\dot{r}_n \approx \ln(2)\,2^{A_n}\dot{A}_n$ which will be used to estimate 
$\dot{A}_n$. Then we insert $\dot{A}_n$ into the transformations of 
Eq.\ref{eq.Local-Var} getting $\dot{E}_n$ and $\dot{C}_n$ as a function of 
$\dot{A}_{n+j},\  j=\{-2,1-,1,2\}$.
To proceed we note that because of the regular dynamics during oscillations 
all the local variables for the different shells are roughly equal despite
a Kolmogorov-scaling of the mean values ($A_n$). This makes us assume 
the following condition between the local variables:
\BE
\left( A_{n+j},E_{n+j},C_{n+j} \right) \approx 
\left( A_n - \mbox{$j \over 3$},E_n,C_n \right),\ j=\{-2,1-,1,2\}
\label{eq.LS=LV}
\EE
When inserted into the different $\dot{A}_{n+j}$ it causes $\dot{E}_n$ and 
$\dot{C}_n$ to resemble $\dot{A}_n$ within pre-factors in front of the 
coupling- and viscous-terms.\\ 
{\em As result the monotony of $\dot{E}_n$ and $\dot{C}_n$ follows that of 
$\dot{A}_n$}.\\
\indent
Now the general flow in $\vec{V}_n$ only depends on the sign of $\dot{r}_n$, 
changing at the manifold and indicated by the arrows showed in 
Fig.\ref{fig.LV-Traj+Surf2}.
From the orientation of the flow and the position of the manifold the 
trajectory is caused to close in on the manifold and slowly drift downwards 
creating a relaxing period.
\subsection{Bursts}
The stability of the manifold and thereby of the relaxing state depends 
critically on the condition of Eq.\ref{eq.LS=LV} used in the derivation above.
The thing that destroys this condition is the approach of a burst from the 
lower shells, affecting only $r_{n-2},r_{n-1}$. The manifold then loses its
stability and the state is forced into a region of strong positive $\dot{r}_n$
making the shell participate in the burst. Now as $r_n$ changes violently it
causes the manifold of the higher shells to become unstable etc.~and thus 
{\em the burst spreads through the shells because of a chain-reaction}.
\section{Conclusion}
In this article the standard GOY shell model has been analysed on the basis of 
its dynamics rather than its statistics. A detailed analysis of the 
time-evolution revels the following:\\
\indent
The dynamics of the model follows two different states where violent bursts 
are interchanged by an oscillatory relaxing state. It is showed that
the dynamics of the shells are mutually correlated and the burst travels
through the shells like a front.
Because bursts in the model cascade nearly unaffected through the shells
in the inertial range, each set of neighbouring shells entering the 
coupling-terms can be seen as local phase-spaces of the corresponding shells,
and when expressed in a simple ``Taylor-series'' base their dynamics describe
an approximative attractor of the model.
From the dynamics of the local attractor the intermittency of the model is 
explained. 
\subsection{Acknowledgements}
I would like to thank the following people for fruitful discussions concerning 
this work: Ken Haste Andersen, Jacob Sparre Andersen, Tomas Bohr, Jesper Borg,
Paolo Muratore Ginanneschi, Martin van Hecke, Anders Johansen, Jens Juul 
Rasmussen, Bjarne Stenum and my supervisor Mogens H{\o}gh Jensen. 
%
%---------------------------------------------------------

%---------------------------------------------------------
\end{document}